\documentclass[conference]{IEEEtran}
\usepackage{amsmath, amssymb, bm, cite, epsfig, psfrag}
\usepackage{graphicx}
\usepackage{soul} 
\usepackage[left=0.625in, right=0.625in, top=0.7in, bottom=0.95in]{geometry}
\usepackage{dblfloatfix}
\usepackage{array}
\usepackage{lipsum}
\usepackage{subcaption}

\newcolumntype{P}[1]{>{\centering\hspace{0pt}}p{#1}}
\newcolumntype{M}[1]{>{\centering\hspace{0pt}}m{#1}}
\newcolumntype{L}{>{\centering\arraybackslash}m{3cm}}
\usepackage{caption}
\usepackage{epstopdf}	
\usepackage{longtable}
\usepackage{supertabular,booktabs}
\usepackage{bbm}
\usepackage{multirow}
\usepackage[usenames,dvipsnames]{xcolor}

\usepackage{etoolbox}
\usepackage{pbox}
\usepackage{fixltx2e}
\usepackage{tabu}	
\usepackage{filecontents}
\usepackage{enumerate}
\usepackage{textcomp}
\usepackage{makecell}
\usepackage{gensymb}
\usepackage{colortbl}
\usepackage{fancyhdr}

\newtoggle{conference}

\togglefalse{conference} 
\graphicspath{{figures/}}

\newcolumntype{?}{!{\vrule width 2pt}}

\fancypagestyle{firststyle}{
	\fancyhf{}
	\fancyhead[L]{S. Ju and T. S. Rappaport, ``Sub-Terahertz Spatial Statistical MIMO Channel Model for Urban Microcells at 142 GHz,''  2021 \textit{IEEE Global Communications Conference (GLOBECOM)}, Dec. 2021, pp. 1-6.}     

}
\IEEEoverridecommandlockouts

\begin{document}
\title{Sub-Terahertz Spatial Statistical MIMO Channel Model for Urban Microcells at 142 GHz} 
\author{\IEEEauthorblockN{Shihao Ju and Theodore S. Rappaport}

\IEEEauthorblockA{	\small NYU WIRELESS, Tandon School of Engineering, New York University, Brooklyn, NY, 11201\\
				\{shao, tsr\}@nyu.edu}
					\thanks{This research is supported by the NYU WIRELESS Industrial Affiliates Program and National Science Foundation (NSF) Research Grants: 1909206 and 2037845.}
}

\maketitle
\thispagestyle{firststyle}
\begin{abstract}
\textcolor{black}{Sixth generation (6G) cellular systems are expected to extend the operational range to sub-Terahertz (THz) frequencies between 100 and 300 GHz due to the broad unexploited spectrum therein. A proper channel model is needed to accurately describe spatial and temporal channel characteristics and faithfully create channel impulse responses at sub-THz frequencies. This paper studies the channel spatial statistics such as the number of spatial clusters and cluster power distribution based on recent radio propagation measurements conducted at 142 GHz in an urban microcell (UMi) scenario. For the 28 measured locations, we observe one to four spatial clusters at most locations. A detailed spatial statistical multiple input multiple output (MIMO) channel generation procedure is introduced based on the derived empirical channel statistics. We find that beamforming provides better spectral efficiency than spatial multiplexing in the LOS scenario due to the boresight path, and two spatial streams usually offer the highest spectral efficiency at most NLOS locations due to the limited number of spatial clusters.} 
\end{abstract}
    

\section{Introduction}
As millimeter-wave wireless systems have been gradually deployed for the fifth-generation (5G) communications, Terahertz (THz) frequencies are expected to play an essential part in the sixth-generation (6G) and beyond, which can provide a tenfold increase in bandwidth \cite{Rap19access}. Sub-THz frequencies (100 GHz - 300 GHz) have been considered the next frontier leading to THz communications over the past few years \cite{Ju21jsac}. Some early works on channel measurements and modeling at 140 GHz have been presented in the literature \cite{Xing21b, Chen21arxiv, Xia20SPAWC, Abbasi20WCNC}. In addition, recent advances in phased arrays, \textcolor{black}{complementary metal–oxide–semiconductor (CMOS)-based power amplifiers}, and low noise amplifiers at sub-THz frequencies \cite{Yamazaki20IET, Xu20IWAT, Hamani20IMS} show that sub-THz wireless systems may become a reality in the near future.

As the carrier frequency increases, the free space path loss increases quadratically according to the Friis' law under the assumption of constant antenna gains \cite{Rap02textbook}. However, the decreasing wavelength at sub-THz frequencies enables tightly packing a larger number of antenna elements into a small form factor. If the physical size of the antenna array (effective aperture) is kept constant at both transmitter (TX) and receiver (RX), theoretically, a quadratic decrease in the free space path loss is expected with increasing frequencies \cite{Xing18gc}. 

Antenna arrays with many antenna elements also bring challenges in many aspects such as signal processing and array modeling \cite{Lu14JSTSP,Ayach14TWC,Sun18twc,Yu16JSTSP}. On the one hand, tightly packed arrays lead to high correlation and mutual coupling across antenna elements \cite{Gao16TAP}. On the other hand, fully digital processing adopted in conventional multiple input multiple output (MIMO) systems at sub-6 GHz may not be realistic for large arrays at sub-THz frequencies since one radio frequency (RF) chain per antenna element will cause prohibitively high power consumption, hardware complexity, and cost \cite{Ayach14TWC}. Thus, hybrid beamforming was proposed by combining the baseband digital processing with the RF analog processing using phase shifters \cite{Sun18twc, Yu16JSTSP, Song16GC}. Even though phase shifters induce constant modulus constraints (i.e., the amplitude of signals cannot be changed) \cite{Sun18twc}, the low cost and near-optimal performance make hybrid beamforming a promising solution for the 5G and beyond \cite{Yu16JSTSP}. Such hybrid architecture employs a few RF chains at the baseband precoding stage and a large number of phase shifters (one or two orders of quantity more than the number of RF chains) to realize spatial multiplexing and beamforming simultaneously.


This paper aims to understand the spatial channel statistics and to evaluate the spectral efficiency of beamforming and spatial multiplexing schemes of sub-THz signals. First, we present the propagation channel measurements conducted at 142 GHz in an urban microcell (UMi) scenario in Section \ref{sec:meas}. The statistics of channel spatial parameters such as the number of spatial clusters, cluster power, and omnidirectional root mean square (RMS) angular spread are derived in Section \ref{sec:stat}. Section \ref{sec:chann_mod} gives a bi-directional MIMO channel generation procedure. Section \ref{sec:se} statistically generates channel impulse responses by implementing the above channel generation procedure and analyzes the spectral efficiency transmitting different number of data streams for a single user. Finally, concluding remarks are provided in Section \ref{sec:conclusion}.

\section{142 GHz UMi Channel Measurements} \label{sec:meas}
\begin{figure}[h!]
	\centering
	\includegraphics[width=1\linewidth]{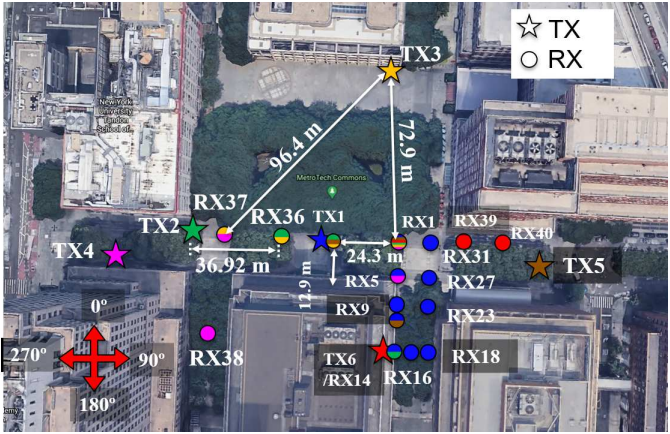}
	\caption{142 GHz outdoor UMi measurement locations \cite{Xing21icc}. TX locations are denoted as stars in different colors. RX locations are denoted as circles. One RX location filled with multiple colors indicates that the channels between this RX location and multiple TX locations were measured. }
	\label{fig:map}
\end{figure}

The 142 GHz UMi channel measurements were conducted in the fall of 2020 on the NYU engineering campus in Downtown Brooklyn, NY, to gain knowledge of potential performance of beamforming and spatial multiplexing of sub-THz wireless systems. As shown in Fig. \ref{fig:map}, the engineering campus includes a square with a size of $\sim$100 m by 100 m, which is surrounded by tall concrete buildings and narrow street canyons, representing a regular UMi scenario. There are glass widows, trees, pedestrians, metal lampposts, and concrete pillars in the surrounding environment causing reflections and scattering. 

A wide range of TX and RX locations were selected for various propagation distances and environments (line-of-sight [LOS] and non-LOS [NLOS]) in order to emulate small cells for sub-THz wireless systems. The TX was deployed at a height of 4 m above ground level serving as a small cell base station (BS), and the RX was placed at a height of 1.5 m above ground level serving as a mobile user (UE). The TX and RX location pairs are provided in Table \ref{tab:TR_pairs}. Overall, there are 16 LOS location pairs and 12 NLOS location pairs with TX-RX (T-R) separation distances from 12.9 m to 117.4 m. 

\begin{figure}[h!]
	\centering
	\includegraphics[width=.8\linewidth]{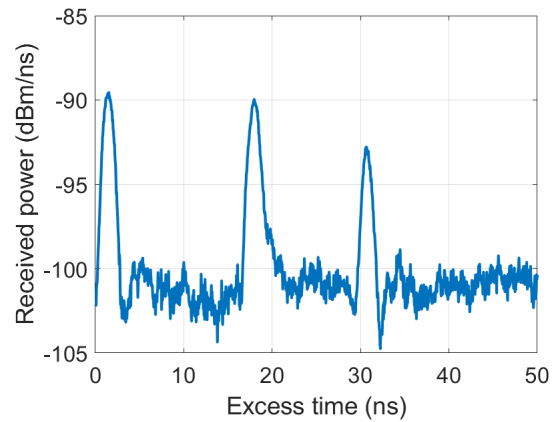}
	\caption{The directional excess time PDP was measured at TX1 and RX14. The received power was recovered by removing system processing gain and antenna gain.}
	\label{fig:pdp}
\end{figure}

Highly directional horn antennas with 27 dBi gain and 8\degree~half-power beamwidth (HPBW) were used at both TX and RX to compensate for the large free space path loss within the first-meter propagation at 142 GHz. The horn antennas and the up/down-converter modules were mounted on electrically controllable gimbals, rotating in the azimuth and elevation planes with a sub-degree resolution. A 500 MHz 11th order pseudorandom m-sequence was generated at the TX baseband and mixed with an intermediate frequency of 7 GHz, then up-converted to a center frequency of 142 GHz. A sliding correlation based receiver was used at the RX, which can provide 39 dB processing gain. The null-to-null RF bandwidth is 1 GHz, which corresponds to a temporal resolution of 2 ns for multipath components (MPCs) \cite{Xing18gc}. \textcolor{black}{The maximum measurable path loss was 152 dB with a minimum signal-to-noise (SNR) requirement of 5 dB.}
\begin{table}[]
	\centering
	\vspace{10pt}
	\caption{\textsc{Measured TX and RX location pairs}}
	\label{tab:TR_pairs}
	\begin{tabular}{|c|c|c|}
		\hline
		\textbf{TX} & \textbf{LOS RX}            & \textbf{NLOS RX}      \\ \hline
		TX1         & RX1, RX5, RX23, RX27, RX31 & RX9, RX14, RX16, RX18 \\ \hline
		TX2         & RX1, RX35, RX36            & RX14                  \\ \hline
		TX3         & RX35, RX36, RX37           & RX1             \\ \hline
		TX4         & RX3, RX37                  & RX1, RX38             \\ \hline
		TX5         & RX3, RX35                  & RX1, RX10             \\ \hline
		TX6         & RX1                        & RX39, RX40            \\ \hline
	\end{tabular}
\end{table}
At each TX and RX location pair, the best TX and RX pointing angles in which the RX received the strongest power were first found and set as the starting points of TX and RX, respectively. The LOS measurements have the TX and RX antennas on boresight, and the NLOS measurements have the TX and RX antennas pointing to the strongest receiving direction by manually searching over azimuth and elevation planes. Then, we fixed the TX pointing angle and swept the RX in the azimuth plane in a step size of the antenna HPBW (8\degree). Each azimuth sweep performed 45 stepped-rotations (360/8=45) and measured 45 directional PDPs. We then performed two identical azimuth sweeps in different elevation planes by uptilting and downtilting the RX by 8\degree. Overall, three RX sweeps produced 135 directional PDPs for one TX pointing angle. The TX was then pointed to some manually selected directions in which the RX detected signal energy above the noise floor. The identical three RX azimuthal sweeps were performed for each individual TX pointing angle. Most TX and RX location pairs have one to three distinct angles of departure (AOD) and angles of arrival (AOA). \textcolor{black}{A sample measured directional PDP having three MPCs is shown in Fig. \ref{fig:pdp}. Note that most measured directional PDPs have only one MPC due to the narrow beamwidths of the TX and RX antennas.}

\begin{figure*}[]
	\centering
	\begin{subfigure}[b]{0.32\textwidth}
		\centering
		\includegraphics[width=\textwidth]{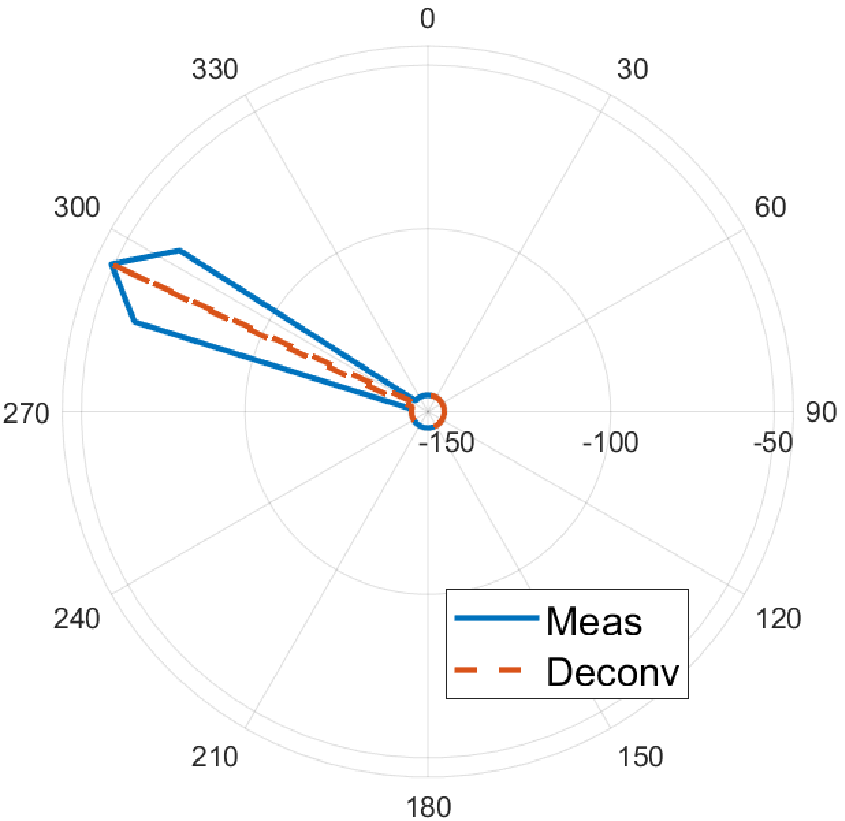}
		\caption{LOS AOA APS: TX1 and RX5}
		\label{fig:aps1}
\end{subfigure}
\vspace{4pt}
	\begin{subfigure}[b]{0.32\textwidth}
	\centering
	\includegraphics[width=\textwidth]{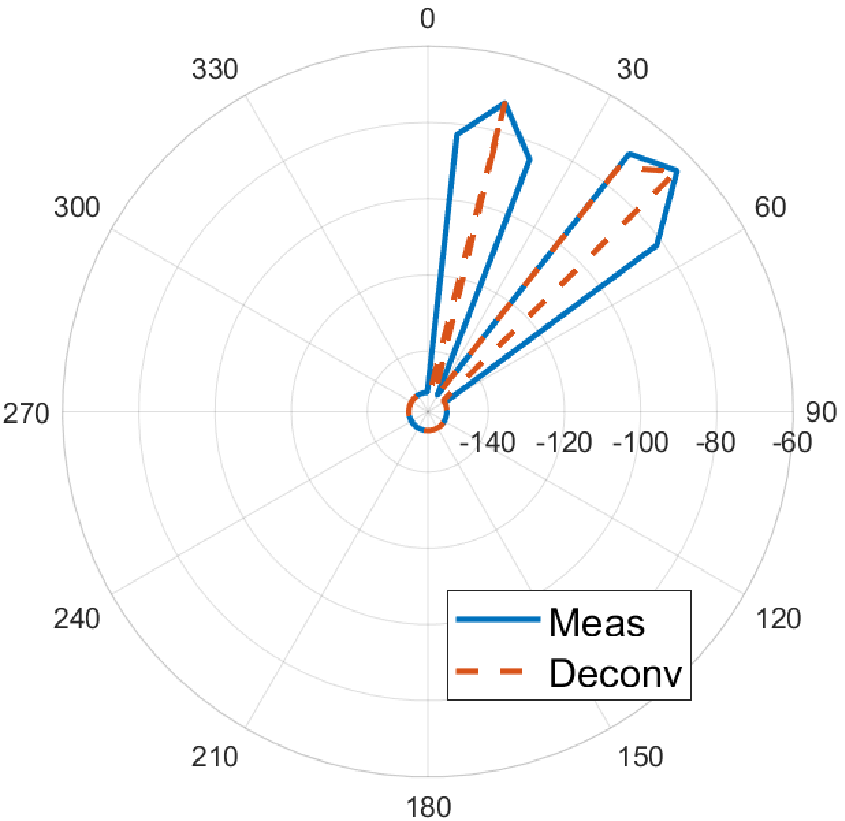}
	\caption{LOS AOA APS: TX3 and RX35}
	\label{fig:aps2}
\end{subfigure}

	\begin{subfigure}[b]{0.32\textwidth}
	\centering
	\includegraphics[width=\textwidth]{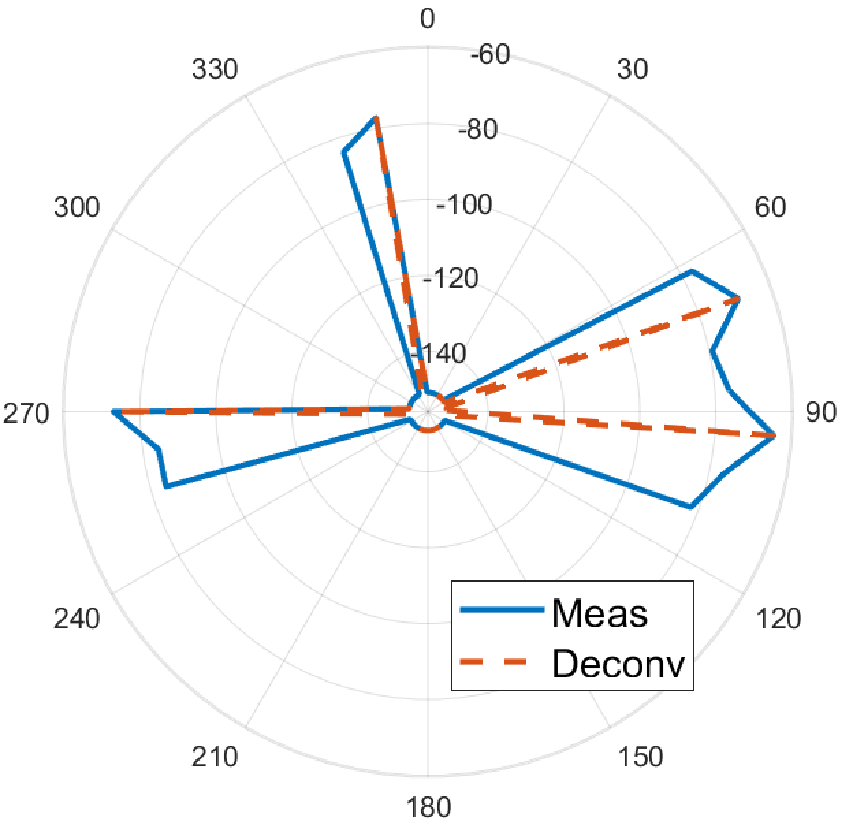}
	\caption{NLOS AOA APS: TX5 and RX1}
	\label{fig:aps3}
\end{subfigure}
\vspace{4pt}
	\begin{subfigure}[b]{0.32\textwidth}
	\centering
	\includegraphics[width=\textwidth]{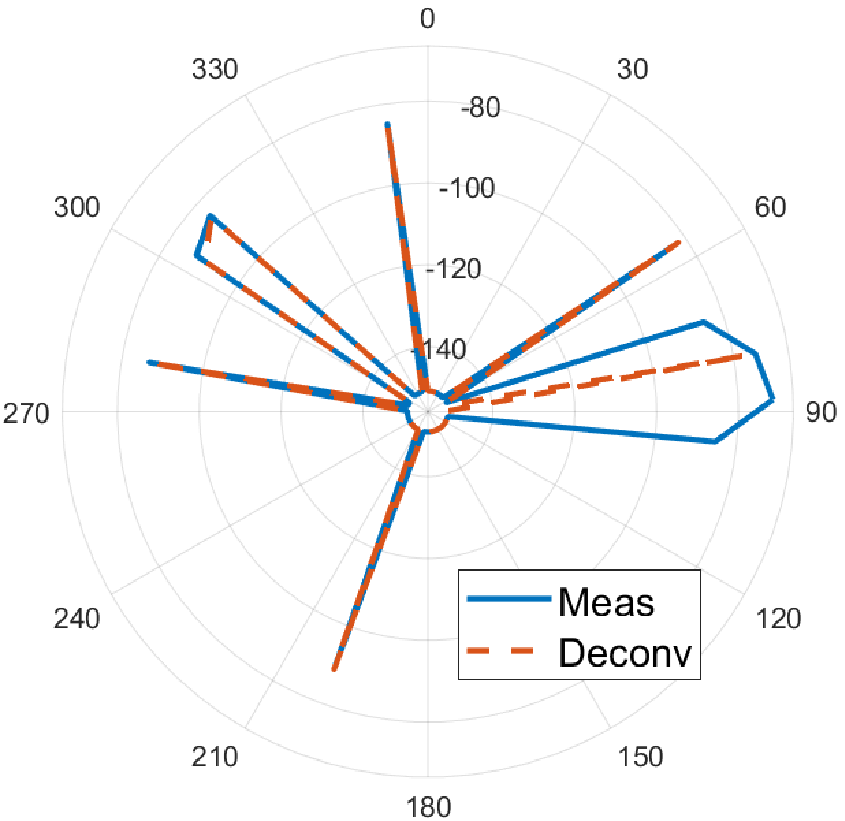}
	\caption{NLOS AOA APS: TX5 and RX10}
	\label{fig:aps4}
\end{subfigure} 
	\caption{2-D AOA angular power spectrum plots with different number of spatial clusters in the boresight elevation angle with a threshold of 20 dB in the directional received power.}
	\label{fig:aps}
\end{figure*}

\section{Channel Spatial Statistics at 142 GHz} \label{sec:stat}
Measurements with narrowbeam horn antennas provide a convenient way to understand channel spatial statistics for sub-THz frequencies. The angular resolution of measurements is limited by the antenna HPBW (8\degree~in this measurements). Even though the azimuth sweeps were performed at three elevation angles, we did not observe two distinct MPCs arriving in a similar azimuth angle but different elevation angles. Thus, we extracted spatial statistics only using the directional measurements conducted in the boresight elevation plane.

Four sample AOA angular power spectrums are shown in Fig. \ref{fig:aps}. The threshold of the angular power spectrum is set to be 20 dB down from the strongest directional power in the azimuth plane. Fig. \ref{fig:aps1} shows the angular power spectrum of a LOS location pair (TX1 and RX5), which only has the boresight path. Fig. \ref{fig:aps2} shows the angular power spectrum of another LOS location pair (TX3 and RX35), which has not only the boresight path but a reflected path from a metal lamppost. From Fig. \ref{fig:aps1} and Fig. \ref{fig:aps2}, we can see that the two neighboring directions next to the peak direction have an approximately 12 dB drop in the received power, which can be the side effect of the horn antenna pattern, which motivates us to recover the actual direction of the arriving signals by deconvolving the antenna pattern out of the angular power spectrum. Thus, the radiation pattern of the employed antennas were measured, and the deconvolved angular power spectrum of AOA are also shown in Fig. \ref{fig:aps} as dashed lines. 


Fig. \ref{fig:aps3} plots the angular power spectrum of an NLOS location pair (TX5 and RX1) showing three spatial lobes. A spatial lobe is defined as a main direction of arrival or departure where multipath components can arrive over hundreds of nanoseconds \cite{Ju21jsac, Samimi16mtt}. The spatial lobes are extracted by partitioning the angular power spectrum with a spatial lobe threshold (e.g., 20 dB used in this paper). A set of consecutively measured directions having powers above the spatial lobe threshold forms a spatial lobe. Fig. \ref{fig:aps4} plots the angular power spectrum of another NLOS location pair (TX5 and RX10) showing six spatial lobes, which is the maximum number of spatial lobes we observed during measurements. Note that in Fig. \ref{fig:aps3}, the angular power spectrum has a wide spatial lobe centered at 90\degree~with two peaks, indicating that two signals were received relatively close in space ($\sim$24\degree~apart) forming a single spatial lobe. Such two signals can be recovered by a deconvolution of the antenna pattern or through peak searching. Due to the limitation of the angular resolution, the subpaths within each signal cannot be resolved in space. We call such a signal a spatial cluster in the rest of this paper. Compared to the definition of spatial lobes, the number of spatial clusters should be no less than the number of spatial lobes. The statistics of spatial clusters will be shown as follows and used in the MIMO channel generation in Section \ref{sec:chann_mod}.

\subsection{The number of spatial clusters}
\begin{figure}[h!]
	\centering
	\includegraphics[width=0.8\linewidth]{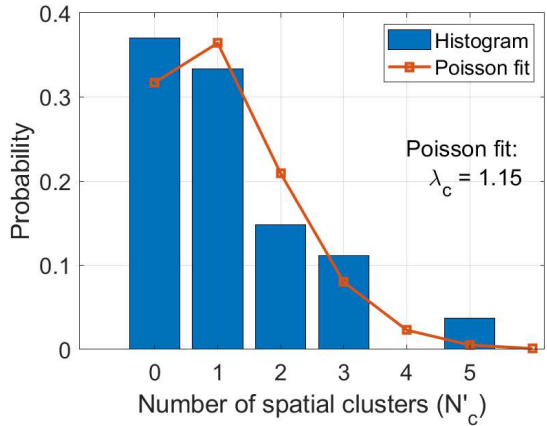}
	\caption{The number of spatial clusters for 142 GHz UMi is well fitted by a Poisson distribution (\ref{eq:poiss}) with $\lambda_c=$ 1.15.}
	\label{fig:num_sc}
\end{figure}

The number of spatial clusters ($N_c$) is inversely proportional to the selected threshold. The histogram of the number of measured spatial clusters over all 28 LOS and NLOS locations in the 142 GHz UMi scenario and the fit Poisson distribution with a mean of 1.15 are shown in Fig. \ref{fig:num_sc}. Since the Poisson starts from zero, and only measurement locations with no outage are considered (i.e., at least one spatial cluster), $N_c'=N_c-1$ is used in the distribution fitting. The Poisson distribution is given by 
\begin{equation}
	\label{eq:poiss}
	\text{Pr}(N_c'=k)=\lambda_c^k \exp(-\lambda_c)/k!, 
\end{equation}
where $\lambda_c$ is the mean of $N_c'$. $k$ can be any integer between zero and five since the maximum number of observed spatial clusters is six. $\lambda_c$ is 0.69 and 1.82 for the separate LOS and NLOS locations, respectively, showing that NLOS locations have more directions of arrival. 

\subsection{Spatial cluster powers}
\begin{figure}[h!]
	\centering
	\includegraphics[width=0.8\linewidth]{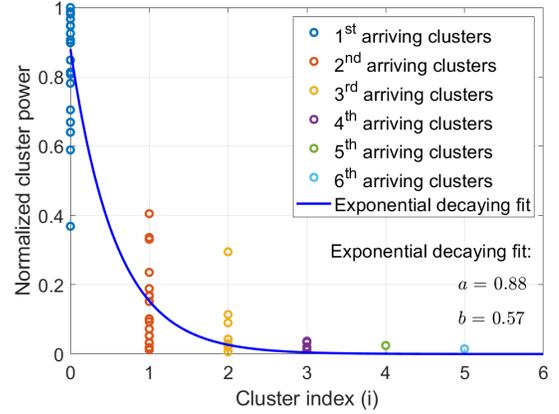}
	\caption{\textcolor{black}{Normalized cluster powers for 142 GHz UMi are well modeled by an exponential decaying function (\ref{eq:cp}) with $a=$ 0.88 and $b=$ 0.57.} }
	\label{fig:sc_powers}
\end{figure}

The received power within each spatial cluster indicates the path quality and determines the practical performance of beamforming and spatial multiplexing implementations. Thus, the distribution of received powers is a key channel indicator for multiple data stream transmission. The normalized cluster powers at each location are calculated by $\bar{P_i}=P_i/\sum_{j=1}^{N_c}P_j$. $P_i$ is the received power in the $i$th cluster, and $\bar{P_i}$ is the normalized cluster power. The order of spatial clusters is determined based on the received power. The first spatial cluster bears the strongest received power. The power distribution among clusters can be modeled as an exponential decaying function of the order of clusters \cite{Ju21jsac}:
\begin{equation}
	\label{eq:cp}
	P_i = ae^{-i/b}10^{c_i/10},
\end{equation}
where $ae^{-i/b}$ represents the average power within the $i$th cluster, and $c_i$ is a shadowing term modeled as a normal random variable with zero mean and a standard deviation of $\sigma_c$ in dB scale. The cluster power distribution at all LOS and NLOS locations and the fit decaying exponential function with $a=$ 0.88 and $b=$ 0.57 are shown in Fig. \ref{fig:sc_powers}. $a$ are 0.96 and 0.76, and $b$ are 0.03 and 0.81 for separate LOS and NLOS scenarios, respectively, showing the single LOS cluster occupies almost all the received power in the LOS scenario whereas the NLOS locations have more clusters bearing comparable powers. 
\subsection{Omnidirectional RMS angular spread}
\begin{figure}[h!]
	\centering
	\includegraphics[width=0.8\linewidth]{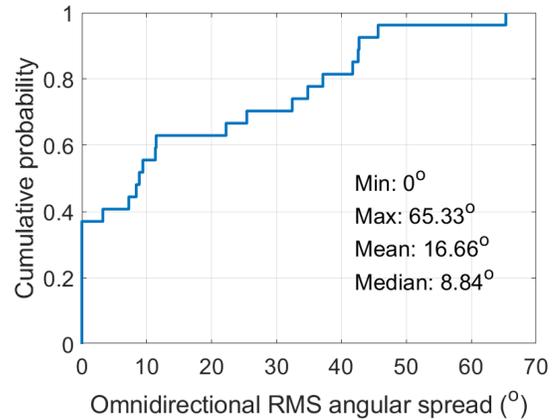}
	\caption{CDF of omnidirectional RMS angular spread.}
	\label{fig:gas}
\end{figure}

Omnidirectional RMS angular spread is an important indicator of angular dispersion of MPCs. Large omnidirectional RMS angular spread suggests that spatial clusters are well separated in the space, which may be beneficial for spatial multiplexing. The empirical cumulative probability function (CDF) of omnidirectional RMS angular spread is shown in Fig. \ref{fig:gas}, where the maximum, mean, and median values are 65.33\degree~, 16.66\degree, and 8.84\degree. 0\degree RMS angular spread suggests that only one spatial cluster exists in the space.

\section{Sub-THz MIMO Channel Generation Procedure} \label{sec:chann_mod}
A double directional MIMO channel generation procedure is presented step by step using the spatial statistics derived in Section \ref{sec:stat}. Due to the lack of relative timing information of signals measured in different directions, a narrowband spatial channel model is assumed here, and a corresponding wideband channel model will be presented in future work. 
\subsubsection{Step 1} Generate the large-scale path loss using the 1 m close-in (CI) free space reference distance path loss model \cite{Xing21b,Ju21jsac,Rap13access}:
\begin{equation}
	\label{eq:pathloss}
		\textup{PL}^{\textup{CI}}(f,d)[\textup{dB}]=\textup{FSPL}(f,1\:\textup{m})+10n\log_{10}(d)+\chi_\sigma, \\
\end{equation}
where $\textup{FSPL}(f,1\:\textup{m}) = 20\log_{10}(4\pi f/c)$. $n$ is path loss exponent (PLE) and $d$ is the 3-D T-R separation distance in meters. Shadow fading $\chi_\sigma$ is modeled by a zero mean Gaussian random variable with standard deviation $\sigma$ in dB. The $n$ and $\sigma$ for 142 GHz outdoor UMi scenario is given in \cite{Xing21icc}, which are 1.94 and 2.66 for the LOS locations, and 2.87 and 8.22 for the NLOS locations. The simulated TX-RX separation distance ranges from 10 to 100 m to agree with the measurements. 

\subsubsection{Step 2} Generate the number of spatial clusters using (\ref{eq:poiss}). Note that the maximum number of generated spatial clusters is limited to six. 
\subsubsection{Step 3} Generate the azimuth AOA and AOD of each spatial cluster. The azimuth plane is equally divided into as many sectors as the number of generated spatial clusters. The central angle of each spatial cluster is uniformly selected in the assigned sector. 
\subsubsection{Step 4} Generate the cluster powers using (\ref{eq:cp}). Note that the generated cluster powers are randomly coupled with the generated cluster angles. 
\subsubsection{Step 5} Generate the phase of each cluster. Because the absolute propagation time of each spatial cluster is unknown, it is reasonable to generate phases following a uniform distribution between 0 and $2\pi$. 
\subsubsection{Step 6} Generate the MIMO channel matrix according to the array geometry:
\begin{equation}
	\textbf{H} = \sum_{i=1}^{N_c}\alpha_i e^{j\psi_i}\textbf{a}_r(\phi_i^r, \theta_i^r)\textbf{a}_t(\phi_i^t, \theta_i^t)^H
\end{equation} 
where $\alpha_i$ and $\psi_i$ denote the channel gain and phase of the $i$th cluster, respectively. $N_t$ and $N_r$ are the number of antenna elements at the TX and RX. $\phi_i^r$ and $\theta_i^r$ are the AOA and ZOA in the global coordinate system (GCS). $\phi_i^t$ and $\theta_i^t$ are the AOD and ZOD in the GCS. $\textbf{a}_t$ and $\textbf{a}_r$ are the spatial signatures of TX and RX antenna arrays, respectively. The superscript $H$ represents the complex conjugate transpose. In this paper, uniform rectangular arrays (URAs) placed in the y-z plane are used at both TX and RX. Such a URA with a size of $N$ elements has $N_h$ elements along the y-axis and $N_v$ elements along the z-axis. The spatial signature is given by \cite{Yu16JSTSP}
\begin{equation}
	\begin{split}
		\textbf{a}(\phi_i,\theta_i) &= \frac{1}{\sqrt{N}}[1, ...,e^{j2\pi\frac{d}{\lambda}(p\sin(\phi_i)\sin(\theta_i)+q\cos(\theta_i))}, \\
		& ..., e^{j2\pi\frac{d}{\lambda}((N_h-1)\sin(\phi_i)\sin(\theta_i)+(N_v-1)\cos(\theta_i))}]^T,
	\end{split}
\end{equation}
where $d$ and $\lambda$ are the antenna space and the signal wavelength. $0\leq p \leq N_h-1$ and $0\leq q \leq N_v-1$ are the indices of antenna elements in the 2-D plane. 

\section{Spectral Efficiency Analysis} \label{sec:se}

Here, we consider a single-cell downlink transmission. Transmit beamforming (precoding) can direct the beam into the direction of the strongest signal using the information of channel matrix obtained via either channel reciprocity or feedback \cite{Sun14CommMag}. Spatial multiplexing enables the TX to transmit different data streams through different MPCs to increase spectral efficiency. Beamforming can be considered a special case of spatial multiplexing with one data stream. Ideally, spatial multiplexing should always have a higher spectral efficiency than beamforming due to the additional spatial degree of freedom \cite{Valenzuela07TWC}. However, several practical constraints make spatial multiplexing unfavorable such as MPC richness and transmit power. Sparse channels at mmWave and sub-THz frequencies in the outdoor environment may not have enough spatial degree of freedom to support high-order spatial multiplexing \cite{Samimi16mtt}. Thus, transmit power needs to be concentrated into only a few strongest directions rather than being split into many beams when the average SNR is low. 
\begin{figure}[]
	\centering
	\begin{subfigure}[b]{0.4\textwidth}
	\centering
	\includegraphics[width=\textwidth]{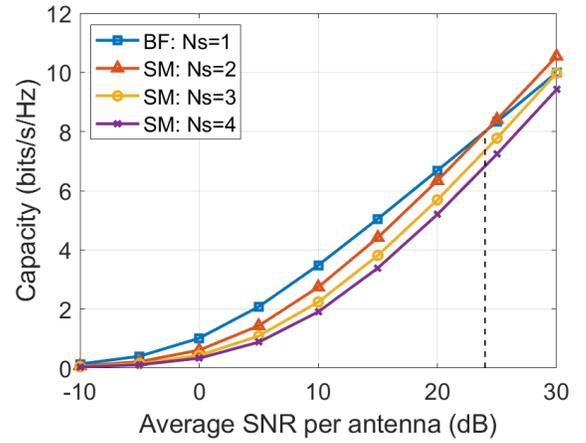}
	\caption{Spectral efficiency in the LOS scenario.}
	\label{fig:se_los}
	\vspace{4pt}
\end{subfigure} 
	\begin{subfigure}[b]{0.4\textwidth}
	\centering
	\includegraphics[width=\textwidth]{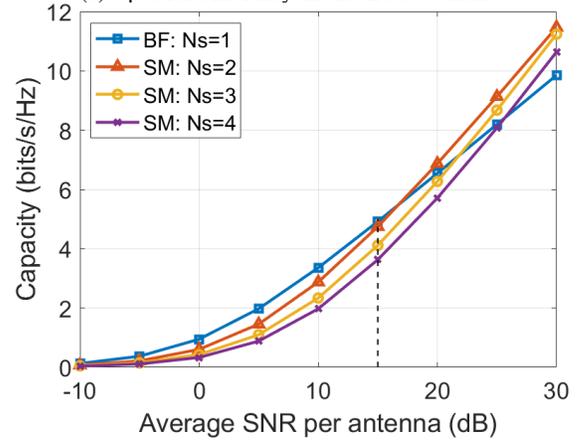}
	\caption{Spectral efficiency in the NLOS scenario.}
	\label{fig:se_nlos}
\end{subfigure} 
	\caption{Spectral efficiency for different number of streams in LOS and NLOS scenarios.}
	\label{fig:se}
\end{figure}

To evaluate the impact of two transmission schemes on the MIMO system throughput, we create thousands of sample channels following the procedure described in Section \ref{sec:chann_mod}. The UEs are randomly dropped in the distance range between 10 and 100 m. A 16 $\times$ 8 URA and a 4 $\times$ 2 URA are deployed at the BS and UE, respectively. Each antenna element is a patch antenna with a length of one wavelength (e.g., 2.11 mm at 142 GHz) and a width of half wavelength, having a maximum directivity of 10.4 dBi and 56\degree~HPBW. The antenna plane is initially set in the y-z plane, and then the broadside is rotated to the direction of the strongest cluster. The number of data streams is set from one to four to investigate the spatial multiplexing gain. The total transmit power is 0 dBm and constant for different number of streams.

We assume that the channel state information is known at the BS, which forms a closed-loop spatial multiplexing. The transmit precoding vector and the receive combining vector are chosen based on the right and left singular vectors of the channel $\textbf{H}$ via the SVD decomposition based on the eigen beamforming (EBF) approach. Fig. \ref{fig:se} shows the spectral efficiencies for different numbers of data streams in the LOS and NLOS scenarios. It should be noted that the beamforming scheme outperforms the spatial multiplexing schemes in the low-SNR region. Spatial multiplexing starts providing additional benefits above a certain SNR threshold, which are 24 dB and 15 dB for the LOS and NLOS scenarios, respectively, suggesting that the LOS path is stronger than the other reflected paths (e.g., 5-15 dB), and the transmit power should be mainly allocated in the LOS direction unless the average SNR in the channel is higher than the threshold. Beamforming should be preferred due to the significant beamforming gain provided by the antenna arrays with massive elements. However, the spatial multiplexing may be much more beneficial in the NLOS environment once the SNR is over 15 dB since the received powers within different spatial clusters are more comparable. \textcolor{black}{In this UMi measurement dataset, we observe two streams can provide higher spectral efficiency than three or four streams at most measured NLOS locations. Even though the maximum number of measured spatial clusters is six, those weak clusters cannot be used for reliable transmission. The cluster power distribution in Fig. \ref{fig:sc_powers} also indicates that the late-arriving clusters bear much less power than the first few arriving clusters.}

\section{Conclusion} \label{sec:conclusion}
\textcolor{black}{This paper presents a spatial statistical MIMO channel model for sub-THz frequencies in the outdoor UMi scenario. 28 TX and RX locations over a wide range of distance from 12.9 m to 117.4 m were selected in an urban open square to conduct propagation measurements at 142 GHz. The number of spatial clusters was modeled as a Poisson random variable, and the average number of spatial clusters is about two over all measurement locations. Thousands of sample channel matrices were simulated using the proposed statistical channel model and used to predict spectral efficiency using either beamforming or spatial multiplexing. Results showed that beamforming may provide higher spectral efficiency when a dominant LOS path exists in the environment. Spatial multiplexing is preferred at most NLOS locations since multiple reflection paths contain comparable signal energy.}


\bibliographystyle{IEEEtran}
\bibliography{gc21}

\end{document}